\documentclass[prper,aps,twocolumn,groupedaddress,floats,showpacs,final,superscriptaddress]{revtex4-1}
\usepackage[latin3]{inputenc}
\usepackage[makeroom]{cancel}
\usepackage{graphicx}
\usepackage{amsmath}
\usepackage{amsfonts}
\usepackage{amssymb}
\usepackage{chngcntr}
\usepackage{color}
\usepackage[left=2cm,right=2cm,top=2cm,bottom=2cm]{geometry}

\usepackage{graphicx} 
\usepackage{dcolumn}
\usepackage{bm}
\usepackage{simplewick}
\usepackage{array}
\usepackage{appendix}

\RequirePackage[
   hyperindex,colorlinks,bookmarksnumbered,
   plainpages=true,pdfstartview=FitH]{hyperref}
\hypersetup{linkcolor=blue,urlcolor=blue,citecolor=blue}
\usepackage{hyperref}
\usepackage{mathrsfs}
\definecolor{purple}{rgb}{0.5,0,0.6}

\usepackage{ulem}
\renewcommand{\emph}[1]{\textit{#1}}
\definecolor{darkblue}{rgb}{0,0,0.5}
\definecolor{darkgreen}{rgb}{0,0.5,0}
\definecolor{darkred}{rgb}{.7,0,0}
\definecolor{purple}{rgb}{0.5,0,0.6}
\definecolor{orange}{rgb}{1,0.5,0}
\definecolor{grey}{rgb}{.6,.6,.6}
\definecolor{lightpink}{rgb}{1,0.7,0.75}
\definecolor{pink}{rgb}{1,0.4,0.58}
\definecolor{deeppink}{rgb}{1,0.08,0.58}

\usepackage{ulem}

\begin{document}

\date{\today}
\title{
 Thermoelectrics of a two-channel charge Kondo circuit: \\
Role of electron-electron interactions in a quantum point contact
}

\author{A. V. Parafilo}
\email{aparafil@ibs.re.kr}
\affiliation{Center for Theoretical Physics of Complex Systems, Institute for Basic Science, Expo-ro, 55, Yuseong-gu, Daejeon 34126, Republic of Korea}

\author{T. K. T. Nguyen}
\affiliation{Institute of Physics, Vietnam Academy of Science and Technology, 10 Dao Tan, Hanoi, Vietnam}

\author{M. N. Kiselev}
\affiliation{The Abdus Salam International Centre for Theoretical Physics, Strada Costiera 11, I-34151, Trieste, Italy}

\date{\today}

\begin{abstract}
In this {\it Letter} we investigate the properties of a quantum impurity model
in the presence of additional many-body interactions between mobile carriers. 
The fundamental question which is addressed here is how the interactions in the charge and spin sectors 
of an itinerant system affect the quantum impurity physics in the vicinity of the intermediate coupling fixed point.
To illustrate the general answer  to this question  we discuss  a two-channel charge Kondo circuit model. 
We show that the electron-electron interactions resulting in the formation
of a massive spin mode in an itinerant electron subset drive the system away from the unstable non-Fermi liquid (NFL) fixed point to the stable Fermi liquid (FL) regime. We discuss the thermoelectric response
as a benchmark for the NFL-FL crossover. 
\end{abstract}

\maketitle

\textit{Introduction.}  The quantum thermoelectricity of
low dimensional systems is a rapidly developing direction of modern
condensed matter physics. Thanks to the incredible development of the
nanotechnology, the fabrication of highly controllable and fine-tunable
nano-devices gives access to a broad variety of charge, spin,
and heat quantum transport phenomena. Since the early 1990s \cite{Streda,Molenkamp,Beenakker,Staring,Humphrey},
thermoelectric efficiency has been predicted to be enhanced in
low dimensional systems in comparison with bulk materials. Moreover,
heat quantization \cite{heatquantization0,heatquantization,heatquantization2},
heat Coulomb blockade \cite{heatblockade}, and the universality of thermoconductance
fluctuations \cite{heatfluctuations} have been investigated in different
nanostructures.

One of the most prominent nano-devices is \textit{a single-electron
transistor} (SET) \cite{devoret}, whose transport properties are
fully governed by the Coulomb blockade  (CB) phenomenon
\cite{shekhter,shekhter2,beenakker}. The SET usually consists of a small island
{[} a so-called quantum dot (QD){]}, connected to electron reservoirs
\cite{kastner} by tunnel barriers or by quantum point contacts (QPC)
\cite{qpc}. With its small size, electrostatically tunable properties,
and sensitive thermoelectric response 
SET provides important information about strong electron-electron
interactions, interference effects, and resonance scattering on the
quantum transport \cite{Blanter,kisbook}. Recent
experiments on the thermopower (TP) in QD systems
quantified the role of sequential tunneling and cotunneling on thermoelectric
transport through the SET \cite{Scheibner} and also demonstrated
pronounced nonlinear thermoelectric effects \cite{Svensson}. 
Moreover, fine tuning the coupling between the QD and leads allowed
access to thermoelectric transport through the Kondo - quantum impurity
\cite{Scheibner} . These promoted studies in both experiment
and theory of thermoelectric transport in QDs in the Kondo regime.

The Kondo effect \cite{kondo}, which shows both resonance scattering
and strong interactions \cite{hewson}, has been detected in SET
fine tuned by the gate to odd CB valleys \cite{Gordon_kondoexp}.
The QD behaves as a quantum spin-$1/2$ impurity \cite{Kouwenhoven _revival}
since the strong correlations between it and the conduction electrons
in the reservoirs lead to the removal of the Coulomb blockade, and result
in a nonmonotonic temperature dependence of the conductance at low
temperatures \cite{Wiel2000_kondoexp,glazmanraikh,kondoexp2}. 

While
the conventional Kondo phenomenon is attributed to a spin degree of
freedom of the quantum impurity, the charge Kondo effect deals with
an iso-spin implementation of the charge quantization. The latter
occurs when a large metallic QD in the Coulomb blockade regime is strongly
coupled to one (or several) lead(s) through a (or several) almost fully transmitting single-mode QPC(s)
\cite{flensberg,matveev,furusakimatveev}. This setup is described
by the Flensberg-Matveev-Furusaki model (FMF). In the absence of a
magnetic field, the FMF setup is mapped into a two
channel Kondo (2CK) model: The left and right moving
modes are treated as isospin variables, whereas the spin projection
quantum numbers of electrons serve as different channels \cite{matveev,furusakimatveev,andreevmatveev}.
Very recently, the FMF model has been achieved in breakthrough experiments \cite{pierre2,pierre3}. These experiments mark an important
step in the study of multichannel Kondo (MCK) problems in which
the universality class known as non-Fermi liquid (NFL) behavior
dominates \cite{blandin}. The NFL picture in the FMF model, however,
is extremely sensitive to variations of external parameters.
Since the intermediate coupling NFL fixed point is unstable, the Fermi
liquid (FL) ground state \cite{LeHur2002,thanh2010}
is achieved by applying relevant small perturbations. For instance,
any small but finite external magnetic field applied to the SET results
in channel asymmetry and thus changes the universality class from
the two-channel Kondo to the single-channel Kondo (1CK) regime \cite{LeHur2002,thanh2010}.

The significant difference of the FMF model compared
to previous theoretical models that have been used to explain
the Kondo effect is that the transmission of electrons through QPCs
happens in one dimension (1D), therefore, the Abelian bosonization
technique \cite{FL_LL1,FL_LL2,gogolin,giamarchi} is applied to solve
the problems. Previous works studying the FMF
model \cite{flensberg,matveev,furusakimatveev,andreevmatveev,LeHur2002,thanh2010}
disregarded the effects of an electron-electron interaction in the QPC(s).
One thus raises an important question regarding whether or not the NFL property can
be broken spontaneously (without any variation of any external parameter)
and how the electron-electron interactions affect
the NFL picture in the FMF model.

In an interacting 1D system, the state resulting from the addition or removal of
an electron may decay quickly into collective charge and spin excitations
which propagate with different velocities (spin-charge separation)
\cite{Haldane}. A theoretical model describing a 1D interacting electron
system which is predicted to behave quite differently from the FL,
is called the Luttinger liquid (LL) model \cite{Tomonaga,Luttinger}.
 The advantage of the LL model is that most of the
interaction processes (namely, the forward $g_{4,||},\:g_{4,\perp},\:g_{2,||},\:g_{2,\perp}$
and the backward $g_{1,||}$ scatterings) \cite{footnoteint} can
be described by the quadratic terms of the bosonic fields. The LL
Hamiltonian in the bosonic representation is thus modified from the
FL one through the effective Fermi velocities $v_{F\rho}$, $v_{F\sigma}$ (the indices $\rho, \sigma$
stand for the charge and spin modes, respectively) and additional dimensionless
parameters $g_{\rho}$, $g_{\sigma}$ (so-called Luttinger parameters) \cite{constants}.
However, the term describing the backscattering $g_{1,\perp}$
process between electrons with opposite spin projection values cannot
be expressed quadratically in bosonic representation, and it must be written explicitly~\cite{giamarchi}. As mentioned in Ref. \cite{giamarchi},
the effects of the $g_{1,\perp}$ term are quite drastic, and it is important
to consider them.

In this Letter we investigate the effects of the electron-electron
interactions in the LL, especially the role of spin-dependent
backward scattering, on the thermoelectric coefficients in the FMF
setup.  In the absence of a $g_{1,\perp}$ process
or in the case when it is irrelevant ($g_{\sigma}\geq1$), the low-temperature scaling behavior of the TP is $S\propto T^{g_{\sigma}-1}\log T$. This scaling is paradigmatic for the two channel Kondo (2CK) model. In addition,
the reflection (transmission) coefficient at the QPC is renormalized due to the electron interactions in the LL
[as known in the Kane-Fisher phenomenon (KFP) \cite{kanefisher}]. Any relevant $g_{1,\perp}$
process appears when $g_{\sigma}<1$ opens a gap in the spin mode (or one says that the
spin field is massive)\cite{luthermass}, and the TP is proportional to the temperature $S\propto T M^{g_{\sigma}/2}$ with $M$ the spin field's mass. We predict that the  backscattering 
process between electrons with opposite spin $g_{1,\perp}$ in the LL can destroy the local NFL-2CK state and drive the
system to the FL-1CK regime. In other words, our results show evidence of the existence of a $g_{1,\perp}$ process if experimentalists find the FL behavior of TP in the FMF setup.

\begin{figure}
\centering \includegraphics[width=1\columnwidth]{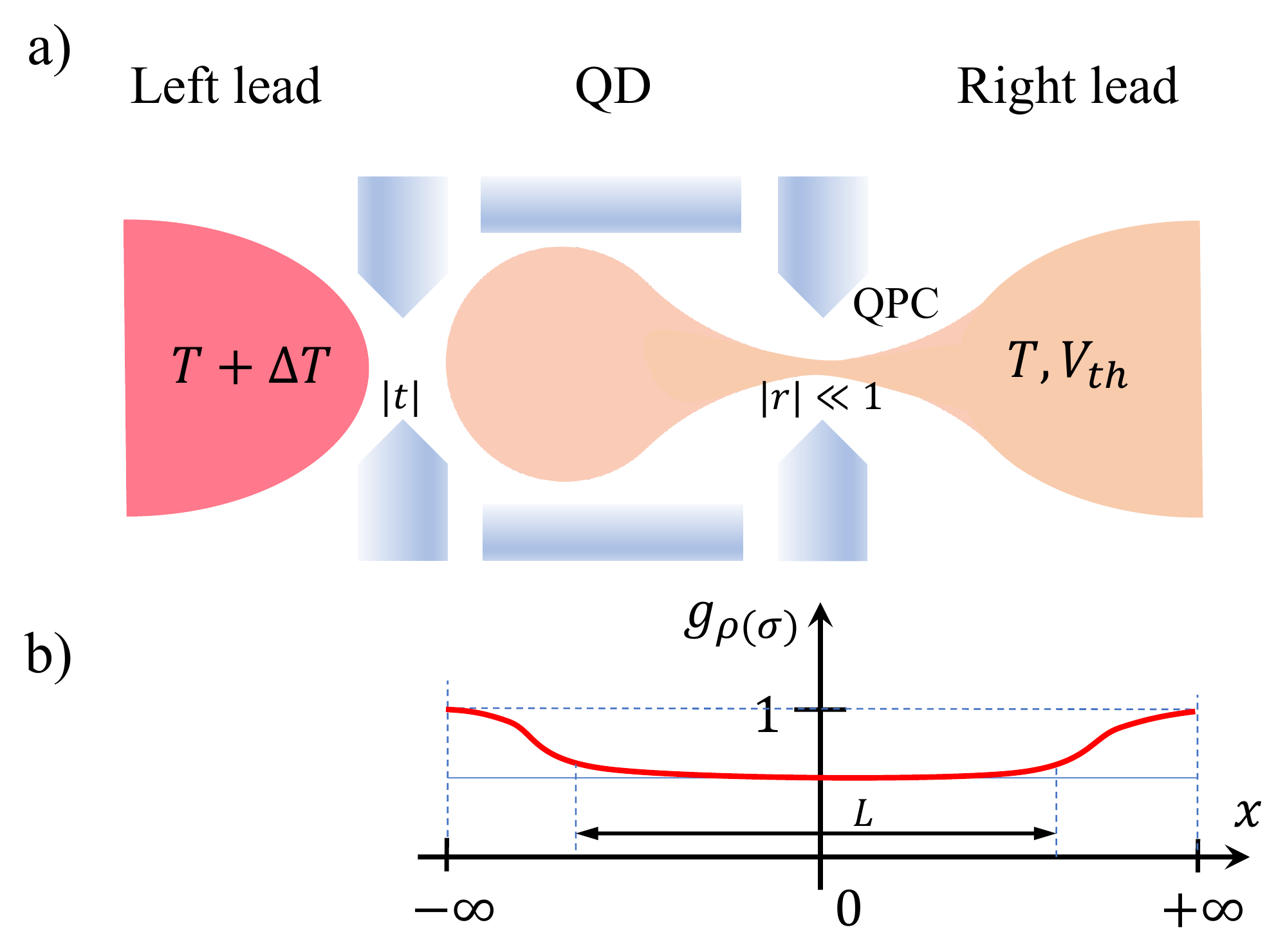} \caption{a) Schematic of a single-electron transistor device: A large
metallic quantum dot (QD) is weakly coupled to the left electrode
through a weak barrier and strongly coupled to the right electrode
through a single-mode quantum point contact (QPC). The QD and electrodes
are formed in a two-dimensional electron gas (2DEG). The tunnel barrier
is characterized by a small transparency $|t|\ll1$, while the electron
scattering in the QPC is determined by a reflection amplitude $|r|\ll1$.
The QD and the right electrode (the drain, marked by the orange
color) are at the reference temperature $T$ while the left electrode
(the source, marked by the red color) is at higher temperature
$T+\Delta T$. A thermovoltage $V_{\rm th}$ is applied to the drain to
compensate the charge flow induced by the temperature
drop $\Delta T$. The effects of electron-electron
interactions in the source are accounted by the Fermi liquid theory.
The role of the electron-electron interaction in the drain, in particular, in the narrow constriction of the size $L$ (the QPC) is the main subject discussed in
this Letter. b) An example showing the evolution of the charge and
spin Luttinger parameters $g_{\rho}$ and $g_{\sigma}$: The interactions
asymptotically vanish both at the position of the tunnel barrier ($x=-\infty$)
and away from the QPC ($x=+\infty$).}
\label{Fig1} 
\end{figure}

\textit{Model.} We consider a nano-device as shown in Fig. \ref{Fig1}
 consisting of a large metallic QD in the weak Coulomb
blockade regime weakly coupled to the left electrode (the source)
via a tunnel barrier and strongly coupled to the right one through
a single-mode QPC. The QD-QPC structure (the drain) is
built of a two-dimensional electron gas (2DEG) and assumed to be in thermal equilibrium at temperature
$T$.
 By applying an external gate, one can control the electron interactions  in the vicinity of the QPC \cite{glazman92, exp_interaction_control1,exp_interaction_control2}. The electrons in this 1D constriction are thus described by the LL model \cite{giamarchi}. The size $L$ 
characterizing the typical length scale for the area where the electron-electron interactions are appreciable is assumed to satisfy the condition that the energy $v_F/L$ is low enough, especially, $v_F/L \ll g_\rho E_C$.
The source separated from the QD by a tunnel contact and considered at
higher temperature $T+\Delta T$ is also formed by 2DEG and can be
described without any loss of generality by conventional Fermi liquid
theory. The temperature drop $\Delta T$ is controlled by using a
current heating technique \cite{Scheibner}. The $\Delta T$ across
the tunnel barrier is assumed to be small compared to the reference
temperature $T$ to guarantee the linear response regime. Applying
thermovoltage $V_{\rm th}$ to implement a zero-current condition for
the electric current between the source and drain allows us to compute the
thermopower (TP) as $S=G_{T}/G|_{I=0}=-V_{\rm th}/\Delta T$, where $G_{T}=I/\Delta T$
is the thermoelectric coefficient, and $G=I/V_{\rm th}$ is the electric conductance.

The weak coupling between the left electrode and the QD is described
by a tunnel Hamiltonian $H_{\rm t}=\sum_{k\alpha}(tc_{k\alpha}^{\dagger}d_{\alpha}+h.c.)$,
where $|t|\ll1$ is the hopping amplitude, and the operators $c_{k\alpha}$ and $d_{\alpha}$
account for electrons with spin ($\alpha=\uparrow,\downarrow$) in the noninteracting left electrode and in
the QD at the point of tunnel junction ($x=-\infty$), respectively.

At the lowest order of perturbation theory over a small transparency
$|t|\ll1$ the transport coefficients can be computed through the
energy dependent tunneling density of states (DOS) related to the
Matsubara Green's function. Here we assume that the DOS of electrons
in the source $\nu_{L}$ is a constant, and electrons in the QD at
the weak link ($x=-\infty$) are noninteracting. We can therefore
apply the Fermi golden rule at the weak link with $\Delta T/T\ll1$.

At the end,
the conductance $G$ and the thermoelectric coefficient $G_{T}$ are
defined through a correlation function $K(\tau)$ of the interacting
electrons in the drain as follows \cite{furusakimatveev,andreevmatveev},
\begin{eqnarray}
 &  & G=G_{L}\frac{\pi T}{2}\int\frac{1}{\cosh^{2}(\pi Tt)}K\left(\frac{1}{2T}+it\right)dt,\label{condcoeff}\\
 &  & G_{T}=-\frac{i\pi^{2}}{2}\frac{G_{L}T}{e}\int\frac{\sinh(\pi Tt)}{\cosh^{3}(\pi Tt)}K\left(\frac{1}{2T}+it\right)dt,\label{thermcoeff}
\end{eqnarray}
where $G_{L}\ll e^{2}/h$ denotes the conductance of the left tunnel
barrier without the influence of the dot.

It is convenient to describe the interacting electrons in the QD-QPC
in the bosonized representation \cite{flensberg,matveev,furusakimatveev,andreevmatveev,LeHur2002,thanh2010}.
In the spirit of Matveev-Andreev theory \cite{andreevmatveev}, the
time-ordered correlation function $K(\tau)$ is computed through the
functional integration over the bosonic fields $\phi_{\uparrow(\downarrow)}(x,t)$, 
\begin{eqnarray}
 &  & K(\tau)=Z(\tau)/Z(0),\label{correlator}\\
 &  & Z(\tau)=\prod_{\alpha=\uparrow,\downarrow}\int\mathcal{D}\phi_{\alpha}\exp\left[-\mathcal{S}_{0}-\mathcal{S}_{C}(\tau)-\mathcal{S}'\right],\label{correlationfunction}
\end{eqnarray}
where $\mathcal{S}_{0}$, $\mathcal{S}_{C}$, and $\mathcal{S}'$ are
Euclidean actions describing the free Luttinger liquid, the Coulomb blockade
in the QD, and the backscattering at the QPC, respectively. The action
$\mathcal{S}_{0}$ is presented as a sum of two independent
actions \cite{FL_LL1,FL_LL2,giamarchi} $\mathcal{S}_{0}=\mathcal{S}_{0}^{(\rho)}+\mathcal{S}_{0}^{(\sigma)}$,
where
\begin{eqnarray}
\mathcal{S}_{0}^{(\rho)} &  & =\frac{v_{F\rho}}{2\pi g_{\rho}}\int dx\int_{0}^{\beta}dt\left[\frac{(\partial_{t}\phi_{\rho})^{2}}{v_{F\rho}^{2}}+(\partial_{x}\phi_{\rho})^{2}\right],\label{chargeaction}
\end{eqnarray}
\begin{eqnarray}
\mathcal{S}_{0}^{(\sigma)} &  & =\int dx\int_{0}^{\beta}dt\left\{ \frac{v_{F\sigma}}{2\pi g_{\sigma}}\left[\frac{(\partial_{t}\phi_{\sigma})^{2}}{v_{F\sigma}^{2}}+(\partial_{x}\phi_{\sigma})^{2}\right]\right.\nonumber \\
 &  & \left.+\frac{2g_{1\perp}D^{2}}{(2\pi v_{F})^{2}}\cos(\sqrt{8}\phi_{\sigma}(x,t))\right\} .\label{spinaction}
\end{eqnarray}
Here, the charge $\phi_{\rho}=(\phi_{\uparrow}+\phi_{\downarrow})/\sqrt{2}$
and spin $\phi_{\sigma}=(\phi_{\uparrow}-\phi_{\downarrow})/\sqrt{2}$
degrees of freedom are separated.{} $v_F$ is the Fermi velocity in the noninteracting system, while $v_{F\rho}$
and $v_{F\sigma}$ are the interaction renormalized Fermi velocities
of charge and spin modes \cite{constants}. The dimensionless charge and spin Luttinger
parameters $g_{\rho}$ and $g_{\sigma}$ characterize the $g_{4,||},\:g_{4,\perp},\:g_{2,||},\:g_{2,\perp},\:g_{1,||}$
electron interaction processes \cite{FL_LL1,FL_LL2,giamarchi}. From
the theory of LL, $0$$<$$g_{\rho(\sigma)}$$<$$1$ $(g_{\rho(\sigma)}$$>$$1)$
describes 1D electrons with a repulsive (attractive) Coulomb interaction,
and $g_{\rho(\sigma)}=1$ corresponds to the non-interacting situation.
The prefactor $g_{1\perp}$ in formula (\ref{spinaction}) characterizes
the $2k_{F}$ spin-flip backscattering in which the fermion fields
with opposite spin projection values are coupled and they exchange
sides of the Fermi surface after the interaction. Due to the fact that
the $g_{1\perp}$ process is not quadratic in a bosonic
representation, its effect is not included in the Luttinger parameters
[see the third (cosine) term in Eq.(\ref{spinaction})].
Therefore, the free action of the spin mode contains a massive term,
in contrast to the massless charge excitation as shown in Eq. (\ref{chargeaction}).
The relevance of the mass can be studied through a \textit{renormalization
group} (RG) analysis of the sine-Gordon model (see, e.g., \cite{gogolin,giamarchi}
for the details). The mass term of the spin mode is irrelevant if $g_{\sigma}\geq1$,
while it is relevant for $g_{\sigma}<1$. In the above equations,
$D$ is an ultra-violet cut-off, which is related to the length parameter
 $a = v_{F}/D$ in the LL-related literature \cite{gogolin,giamarchi}, and $\beta=1/T$
(here we adopt units $\hbar=k_{B}=1$).

The Coulomb interaction in the QD is described by the Hamiltonian
$H_{C}=E_{C}[\hat{n}-N]^{2}$, where $E_{C}=e^{2}/2C$ is the charging
energy ($C$ is the QD capacitance), and $\hat{n}$$=$$\hat{n}_{L}$$+$$\sum_{\alpha=\uparrow,\downarrow}\phi_{\alpha}(0,t)/\pi$
is the operator of the number of electrons entered through the tunnel
barrier and the QPC, respectively \cite{aleiner_glazman}; $N$ is
a dimensionless parameter proportional to the gate voltage $V_{g}$.
Without loss of generality, the number of electrons entering the dot
from the left electrode can be replaced by a time-dependent function
$n_{\tau}=\theta(t)\theta(\tau-t)$, where $\theta(t)$ is the Heaviside
step-function. Therefore, the Coulomb blockade action $\mathcal{S}_{C}$
in bosonic representation reads \cite{flensberg,matveev,furusakimatveev,andreevmatveev,LeHur2002,thanh2010,aleiner_glazman}
\begin{eqnarray}
\mathcal{S}_{C} & = & E_{C}\int_{0}^{\beta}dt[n_{\tau}(t)+\frac{\sqrt{2}}{\pi}\phi_{\rho}(0,t)-N]^{2}.
\end{eqnarray}

Finally, the contribution $\mathcal{S}'$ in the action of the QD-QPC
structure characterizes the weak backscattering at the QPC, 
\begin{eqnarray}
\mathcal{S}'\! & = & -\frac{2D}{\pi}|r|\!\int_{0}^{\beta}\!\!\!dt\cos[\sqrt{2}\phi_{\rho}(0,t)]\cos[\sqrt{2}\phi_{\sigma}(0,t)],\label{eq:BS}
\end{eqnarray}
where $|r|\ll1$ is a small reflection amplitude. Interestingly,
one notices that both the $g_{1\perp}$ interaction process in the
LL and the backscattering (\ref{eq:BS}) happen simultaneously at
the QPC.

\textit{Massless spin field.} We first study the situation in which
the spin field $\phi_{\sigma}$ is massless. In
accordance with the RG analysis \cite{giamarchi} it occurs when $g_{\sigma}\geq1$.

In the absence of backscattering $r=0$, the functional integral Eq.(\ref{correlationfunction})
is Gaussian. The correlator $K_{0}(\tau)\equiv K(\tau)|_{r=0}$ is
computed at low temperature $T\ll E_{C}$ and at $\tau\gg E_{C}^{-1}$.
The main contribution to the electric conductance is the zero order
term of the perturbation expression over the reflection amplitude $|r|$
with the condition we will mention later. Therefore, 
\begin{eqnarray}
G & = & G_{L}C(g_{\rho})\left(\frac{T}{g_{\rho}E_{C}}\right)^{\frac{1}{g_{\rho}}},\label{conductance}
\end{eqnarray}
with 
\begin{eqnarray}
C(g_{\rho}) & = & \frac{\sqrt{\pi}}{2}\left(\frac{\pi^{2}}{2\gamma}\right)^{1/g_{\rho}}\frac{\Gamma\left(1+g_{\rho}/2\right)}{\Gamma\left(3/2+g_{\rho}/2\right)}
\end{eqnarray}
depends only on the value of the charge Luttinger parameter $g_{\rho}$,
$\Gamma\left(y\right)$ is the gamma function, $\gamma=e^{C}$, and $C\approx0.577$
is Euler's constant. The electron interactions in the LL renormalize
both the scaling of the conductance ($G\propto T^{1/g_{\rho}}$) and
the charging energy ($g_{\rho}E_{C}$). Note that at $r=0$ the conductance
depends only on the interaction in the charge mode through the parameter
$g_{\rho}$. The integrals over the spin field $\phi_{\sigma}$ are
unaffected by $n_{\tau}\left(t\right)$, and the correlator $K_{0}(\tau)$
is thus independent from the free spin mode action in Eq. (\ref{spinaction}).
If $g_{\rho}=1$, we restore the result for the noninteracting case $G=(\pi^{3}G_{L}/8\gamma)(T/E_{C})$
as shown in Ref. \cite{furusakimatveev}. The temperature scaling
of the conductance in Eq. (\ref{conductance}) is relevant with the
results explained in Refs. \cite{flensberg,matveev,thanhprl,yikane,thanh2020}:
$G\propto T^{2/\mathcal{M}}$, where $\mathcal{M}$ is the number of channels
in the charge Kondo effect (it is two, the number of electron's spin
projection in FMF model or the number of the QPCs in the experimental
integer quantum Hall setup \cite{pierre2,pierre3,thanhprl}). 
In addition, the $G\propto T^{1/g_{\rho}}$ scaling also represents the fact that there are 
 no stable electron-like quasiparticles in the LL. As a consequence,
the quasiparticle residue vanishes and the power-law behavior appears
in many observables \cite{kanefisher,furusakinagaosa1,pierre,furusaki}.

The thermoelectric coefficient $G_{T}$ vanishes at the $|r|=0$ limit
due to the electron-hole symmetry. A finite contribution in $G_{T}$
is computed in perturbation theory over a small reflection coefficient
$|r|\ll1$. Expanding the partition function Eq.~(\ref{correlationfunction})
over $\mathcal{S}'$, we obtain $K(\tau)=K_{0}(\tau)[1+(\langle\mathcal{S}'^{2}\rangle_{\tau}-\langle\mathcal{S}'^{2}\rangle_{0})/2]$.
One should notice that the fluctuations of the massless spin mode
are not suppressed at low energies, and the average $\langle\mathcal{S}'\rangle$
vanishes. Thus, a nonvanishing backscattering correction to the correlation
function appears only in the second order in $|r|$. The thermoelectric
coefficient $G_{T}$ is computed with logarithmic accuracy as 
\begin{eqnarray}
G_{T} & = & -\frac{G_{L}|r^{\ast}|^{2}}{e}A(g_{\rho},g_{\sigma})C_{T}(g_{\rho},g_{\sigma})\sin(2\pi N)\nonumber \\
 &  & \times\log\left(\frac{E_{C}}{T}\right)\left(\frac{T}{g_{\rho}E_{C}}\right)^{\frac{1}{g_{\rho}}+g_{\sigma}-1},\label{thermo-coefficient}
\end{eqnarray}
where $|r^{\ast}|=|r|(g_{\rho}E_{C}/D)^{(g_{\rho}+g_{\sigma})/2-1}$
is the interaction renormalized reflection amplitude \cite{footnote},
and the interaction dependent pre-factors,
$A(g_{\rho},g_{\sigma})=\left(2\gamma\right)^{g_{\rho}-\frac{1}{g_{\rho}}}\pi^{\frac{2}{g_{\rho}}-g_{\rho}+g_{\sigma}-1}$, and
\begin{eqnarray}
C_{T}(g_{\rho},g_{\sigma})=
\int_{-\infty}^{\infty}\!\!\!dz\frac{\sinh(z)}{[\cosh(z)]^{3+\frac{1}{g_{\rho}}}}\left\{ \tilde{F}(z_{-})-\tilde{F}(z_{+})\right\}.\;\;\;\;
\end{eqnarray}
Here, we define $\tilde{F}(z)$$=$$e^{ig_{\sigma}(z-\pi/2-i\ln2)}$
$\times$$\,_{2}F_{1}[g_{\sigma}/2,g_{\sigma},(2+g_{\sigma})/2,e^{2iz}]/g_{\sigma}$, where
$\,_{2}F_{1}$$(a,b,c,d)$ is the hypergeometric function, and $z_{\pm}=\pi/2\pm iz$.
The thermoelectric coefficient $G_{T}$ shows the temperature dependent
$T^{1/g_{\rho}+g_{\sigma}-1}{\rm log}T$ scaling. In the noninteracting
regime, $g_{\rho}$$=$$g_{\sigma}$$=$$1$, it recalls the result
$G_{T}\propto T\log(T)$ in Ref. \cite{andreevmatveev}.

The effect of the electron interaction on $G_{T}$ is threefold: (i)
the power-law temperature dependence $G_{T}\propto(T)^{1/g_{\rho}+g_{\sigma}-1}$;
(ii) the renormalization of the charging energy $E_{C}\rightarrow g_{\rho}E_{C}$;
and (iii) the renormalization of the weak scattering potential at the QPC
$r\rightarrow r^{\ast}$. The last effect reveals the KFP \cite{kanefisher}.
The interaction renormalized reflection amplitude is consistent with
the corresponding RG analysis showing that if the interaction
in the LL ($g_{\rho},g_{\sigma}<1$) is repulsive, the effective reflection
amplitude increases, achieving the weak coupling limit ($r^{\ast}\rightarrow1$).
On the contrary, the scattering at the QPC becomes irrelevant ($r^{\ast}\rightarrow0$)
for the attractive interactions ($g_{\rho},g_{\sigma}>1$). 

Plugging Eqs.~(\ref{conductance}) and (\ref{thermo-coefficient})
into the definition formula of TP, $S=G_{T}/G$, we obtain 
\begin{eqnarray}
\!\!\!S & = & \!\!-\frac{|r^{\ast}|^{2}}{e}C_{S}(g_{\rho},g_{\sigma})\sin(2\pi N)\log\left(\!\!\frac{E_{C}}{T}\!\!\right)\!\!\left(\!\!\frac{T}{g_{\rho}E_{C}}\!\!\right)^{g_{\sigma}-1}\!\!\!\!\!\!,\label{masslessTP}
\end{eqnarray}
where $C_{S}(g_{\rho},g_{\sigma})=A(g_{\rho},g_{\sigma})C_{T}(g_{\rho},g_{\sigma})/C(g_{\rho})$.
The temperature scaling of TP, $T^{g_{\sigma}-1}{\rm log}T$,
in Eq.~(\ref{masslessTP}) vanishes when $T\rightarrow0$ for $g_{\sigma}>1$.
This zero-temperature vanishing characteristic is consistent with
the corresponding non-perturbative scaling of the TP maximum for 2CK
in Ref.~\cite{andreevmatveev}. In the charge Kondo effect the charge
mode is always blockaded locally, while the spin mode usually fluctuates
freely. The gapless spin mode is responsible for the NFL behavior.
Therefore, only the spin Luttinger parameter appears in the $T^{g_{\sigma}-1}{\rm log}T$
scaling. The effect of the interaction in the spin mode becomes more
dominant in the case when the spin mode is massive. In addition, the
TP $S$ in Eq.~(\ref{masslessTP}) diverges at zero temperature in
the noninteracting spin field case $g_{\sigma}=1$, showing the breakdown
of the perturbation theory at sufficiently low temperature. Thus,
the validity the perturbation theory is justified by the condition for
temperature: $|r^{\ast}|^{2}g_{\rho}E_{C}\ll T\ll g_{\rho}E_{C}$ \cite{validity}
 if $|r^{\ast}|^{2}g_{\rho}E_{C} > v_F/L$. 
Another dramatic manifestation of the LL properties in the behavior
of TP is the appearance of the KFP through the renormalized reflection
amplitude at the QPC $|r^{\ast}|$, which is completely different
from the results shown in Refs. \cite{krive1,krive2,krive3,yang}.

 Note that Eqs.~(\ref{conductance}) and (\ref{thermo-coefficient}) can be obtained by applying the spatially inhomogeneous Green's function method for the finite LL wire in the so-called ``high temperature'' regime $v_{F\rho(\sigma)}/L\ll T\ll g_{\rho}E_C$ \cite{maslov, safi, furusakilead}. In fact, the theory of quantum transport in a 2CK -- FMF model with a finite LL wire demonstrating the QPC vicinity is studied \cite{thanh2022}. We find that the temperature scalings of the thermoelectric coefficients are independent of the LL length $L$. For the purpose of investigating the electron interaction effects, we  focus on a discussion of the limit $L\to\infty$. The finite-$L$ effects will be considered elsewhere \cite{thanh2022}. 

\textit{Massive spin field.} In this section, we address the question of how the pinning
potential of the spin mode {[}cosine term in Eq.(\ref{spinaction}){]} affects the thermoelectric properties of the spinful LL-based
QD-QPC structure in the case $g_{\sigma}<1$. The saddle point solution of the spin mode is $\phi_{\sigma,sp}=2\pi n/\sqrt{8}$
for $g_{1\perp}<0$, while $\phi_{\sigma,sp}=\pi/\sqrt{8}+2\pi n/\sqrt{8}$
for $g_{1\perp}>0$ \cite{giamarchi} ($n$ is an integer  number).

The free action of the spin fluctuations $\varphi_{\sigma}=\phi_{\sigma}-2\pi n/\sqrt{8}$
around the saddle point solution reads as
\begin{eqnarray}
\mathcal{S}_{0}^{(\sigma)}=\int dx\int_{0}^{\beta}dt &  & \left\{ \frac{v_{F\sigma}}{2\pi g_{\sigma}}\left[\frac{(\partial_{t}\varphi_{\sigma})^{2}}{v_{F\sigma}^{2}}+(\partial_{x}\varphi_{\sigma})^{2}\right]\right.\nonumber \\
 &  & \left.+\frac{2g_{1\perp}D^{2}}{\pi^{2}v_{F}^{2}}\varphi_{\sigma}^{2}\right\} .\label{massspinaction}
\end{eqnarray}
From Eq.~(\ref{massspinaction}), the mass of the spin mode can be
defined as $M$$=$$2D(v_{F\sigma}/v_F)\sqrt{g_{\sigma}|g_{1\perp}|/\pi v_{F\sigma}}$.
However, the correct value of $M$ should be found self-consistently
by applying Feynman's variational principle \cite{giamarchi} or a more
strict RG analysis of the sine-Gordon model \cite{gogolin,giamarchi}. Both methods give a more
complicated dependence of the mass on the interaction constant and
Luttinger parameter, $M$$=$$D(v_{F\sigma}/v_F)(|g_{1\perp}|/\pi v_{F\sigma})^{1/(2-2g_{\sigma})}$,
in comparison with the perturbative analysis. We will use the latter
expression of $M$ in this Letter.

The spin mode of the LL is now pinned at low energies, and the nonvanishing
backscattering correction to the correlation function $K(\tau)$ is
thus obtained in the first order of the perturbation theory over the
small reflection amplitude ($|r|\ll1$): $K(\tau)=K_{0}(\tau)\left[1-\langle\mathcal{S}'\rangle_{\tau}+\langle\mathcal{S}'\rangle_{0}\right]$.
Straightforward calculations give the expression of the TP at low
temperature $T\ll M, g_{\rho}E_{C}$ as 
\begin{eqnarray}
\!\!\!\!\!\!\!\!\!\!S\! & =\!\! & -\frac{1}{e}|r^{\ast}|C_{S}^{\ast}(g_{\rho})\sin(2\pi N)\frac{T}{g_{\rho}E_{C}}\!\left(\!\frac{M}{2\sqrt{2}g_{\rho}E_{C}}\!\right)^{\frac{g_{\sigma}}{2}}\!\!\!\!\!,\label{massiveTP}
\end{eqnarray}
in which the interaction dependent prefactor is 
\begin{eqnarray}
\!\!\!\!\!\!C_{S}^{\ast}(g_{\rho})\!=\!\frac{\xi\pi^{4+g_{\rho}}}{C(g_{\rho})}\!\!\left(\!\frac{2\gamma}{\pi^{2}}\!\right)\!^{\frac{g_{\rho}}{2}-\frac{1}{g_{\rho}}}\!\!\!\int_{-\infty}^{\infty}\!\!\!\!\!dy\frac{\sinh^{2}(y)}{\cosh^{4+\frac{1}{g_{\rho}}}(y)},
\end{eqnarray}
where $\xi\approx1.59$ \cite{andreevmatveev,yikane}. Similar to
the way that the Coulomb blockade acts on the charge fluctuations,
the existence of the finite mass in the spin mode suppresses its fluctuations
around the saddle-point at low energies, the TP Eq.~(\ref{massiveTP})
is thus proportional to $|r^{\ast}|$ and temperature. The ratio of
two ``masses'' $M/g_{\rho}E_{C}$ determines the strength of TP
at a given spin mode Luttinger parameter $g_{\sigma}$. In fact, the
backscattering at the QPC, which determines the efficiency of the
thermoelectric transport, is similar to the backward
interaction $g_{1\perp}$ process of the LL. Therefore, the influence of
the $g_{1\perp}$ process on the TP is dominant.

Equations~(\ref{masslessTP}) and (\ref{massiveTP}) represent the
central results of this Letter. Interestingly, the TP for the massless
spin mode case depends on temperature nonmonotonically {[}$S\propto T^{g_{\sigma}-1}\log T$,
with $g_{\sigma}\geq1$, as shown in Eq.~(\ref{masslessTP}){]}
while the TP for the massive spin mode case with $g_{\sigma}<1$ is proportional
to the temperature {[}$S\propto T$, as shown in Eq.~(\ref{massiveTP}){]}.
The former shows the NFL property characterizing the 2CK, while the
latter shows the FL picture of the 1CK. What physical quantities
control this crossover from 2CK to 1CK? Notably different from the
fact that a finite \textit{external} magnetic field breaks the symmetry
of the up-spin and down-spin as explained in Ref.~\cite{thanh2010},
our current results show that the relevant backward
$g_{1\perp}$ scattering process in the LL (at $g_{\sigma}<1$) induces the
instant asymmetry of these two Kondo channels at the QPC. This Kondo
channel symmetry breaking induces the crossover from 2CK to 1CK. An
alternative point of view for the 2CK-1CK crossover in this work can
be represented as follows: In the charge Kondo effect the charge mode
is always blockaded (locally) while the spin mode is usually unblockaded.
It refers to the gapless spin mode. This gapless mode results in the
local NFL property of 2CK, in contrast to the local FL appearing in the
1CK regime. If the spin mode is additionally gapped (e.g., either by
a ``trivial'' Zeeman effect or by ``nontrivial'' many-body effects
in the LL), the local NFL state is destroyed.

 It was argued (see, e.g. \cite{Meir})  that interactions in a QPC are the source of the so-called ``0.7-anomaly'' and Kondo-like effects are invoked for those explanations \cite{Meir}, but recent studies show no link between the Kondo effect and the ``0.7-anomaly'' \cite{exp_07anomaly}. In this work, we deal with the mesoscopic Coulomb blockade which assumes a weak charge quantization and does not pronounce conductance steps. However, reducing interactions in the QPC is helpful for the enhancement of TP and the protection of the NFL properties. Therefore, it is necessary to fabricate a clean ballistic QPC \cite{Gordon}.

\textit{Conclusions. } In this Letter, we have  investigated
theoretically the influence of the electron interactions in the
LL based QD-QPC structure on the two-channel charge Kondo problem.
Using the Abelian bosonization technique and calculating the thermoelectric
coefficients perturbatively with respect to the reflection amplitude
at the QPC, we predict the low-temperature scaling behavior of the
Seebeck coefficient as $S\propto T^{g_{\sigma}-1}\log T$ for the
massless spin mode case and $S\propto T$ for the massive
one. We predict that the relevance of the backscattering
$g_{1\perp}$ process 
 induces a universal crossover from NFL-2CK to FL-1CK. It
opens an interesting possibility for investigating the  crossover
between multi- and single-channel Kondo regimes in experiments.

\textbf{Acknowledgements.} A.V.P. thanks I. V. Krive and S. I. Kulinich
for fruitful discussions. A.V.P. acknowledges support by Institute for
Basic Science in Korea (IBS-R024-D1). This research in Hanoi is funded
by Vietnam National Foundation for Science and Technology Development
(NAFOSTED) under grant number 103.01-2020.05. {\color{black} The work
of M.K. is conducted within the framework of the Trieste Institute
for Theoretical Quantum Technologies (TQT).}


\begin{thebibliography}{10}
\bibitem{footnoteint} In 1D systems, we deal with
excitations close to the Fermi surface (FS), so we linearize the dispersion
relation close to each Fermi point $\pm k_{F}.$ The low-energy processes
of the electron-electron interaction can be decomposed in three sectors.
The $g_{4,p}\sim U(0)$ process couples only electrons on the same
side of the FS [$U\left(q\right)$ is the Fourier transform of the
interaction potential], the $g_{2,p}\sim U(0)$ process couples electrons
from one side of the FS with ones on the other side, while the $g_{1,p}\sim U(2k_{F})$
process corresponds to the $2k_{F}$ backscattering where the electrons
exchange sides. The two former processes are forward scattering since
each species of electrons stays on the same side of the FS after the
interactions \cite{giamarchi}. Additional indices $p=\parallel,\perp$
denote whether the spins of two electrons participated in the scattering
are equal or opposite.
\bibitem{constants} The effective Fermi velocity
$v_{F\alpha}$ and the dimensionless Luttinger liquid parameter $g_{\alpha}$
for the charge ($\alpha$$=$$\rho$) and spin ($\alpha$$=$$\sigma$)
modes are defined as $v_{F\alpha}=$$\sqrt{\left(2\pi v_{F}+g_{4\alpha}\right)^{2}-g_{12\alpha}^{2}}/2\pi$, 
$g_{\alpha}=\sqrt{\left(2\pi v_{F}+g_{4\alpha}+g_{12\alpha}\right)/\left(2\pi v_{F}+g_{4\alpha}-g_{12\alpha}\right)}$,
where $v_F$ is the Fermi velocity,  $g_{4,\alpha}=g_{4,\parallel}\pm g_{4,\perp}$ and $g_{12\alpha}=g_{1,\parallel}-g_{2,\parallel}\mp g_{2,\perp}$
\cite{giamarchi}. Here the upper and lower signs stand for $\alpha=\rho,\:\sigma$, respectively.
$v_{F,\alpha}=v_{F}$ and $g_{\alpha}$$ =1$ in the absence of electron-electron
interactions.
\bibitem{footnote} We would like to comment about the appearance of the
ultraviolet cutoff $D$ in all final results among the renormalized
coefficient $|r^{\ast}|$. Putting $D\rightarrow\infty$ will lead to
an ultraviolet divergency of all results. This is an artifact of the LL
model, where the interaction is assumed to be point like. However, taking
into account the finite size of the Coulomb interaction $\tilde a$ will
result in replacing $D/v_{F}\rightarrow \tilde a^{-1}$. Thus, the value
of the renormalized reflection coefficient remains as the fitting
parameter.
\bibitem{validity}
The validity of the perturbation theory for the $G_T$ in the massless spin field case should be determined from Eq.~(\ref{thermo-coefficient}). Thus, the lower bound value for the Luttinger parameter $g_{\rho}^{\rm min}$ is found using the following condition: $|r^{\ast}|^2A(g_{\rho}^{\rm min},g_{\sigma})C_T(g_{\rho}^{\rm min},g_{\sigma})\sim 1$. Assuming $|r|=0.2$ and $E_C/D=0.1$ we estimate numerically $g_{\rho}^{\rm min}\approx 0.51$. An analogous calculation for the massive spin mode case gives $g_{\rho}^{\rm min}\approx 0.63$.
\end{thebibliography}
\end{document}